# Suspension and Measurement of Graphene and Bi$_2$Se$_3$ Atomic Membranes

Jairo Velasco Jr. , Zeng Zhao, Hang Zhang, Fenglin Wang, Zhiyong Wang, Philip Kratz, Lei Jing, Wenzhong Bao, Jing Shi and Chun Ning Lau

Department of Physics and Astronomy, University of California, Riverside, Riverside, CA 92521

**Abstract**

Coupling high quality, suspended atomic membranes to specialized electrodes enables investigation of many novel phenomena, such as spin or Cooper pair transport in these two dimensional systems. However, many electrode materials are not stable in acids that are used to dissolve underlying substrates. Here we present a versatile and powerful multi-level lithographical technique to suspend atomic membranes, which can be applied to the vast majority of substrate, membrane and electrode materials. Using this technique, we fabricated suspended graphene devices with Al electrodes and mobility of 5500 cm$^2$/Vs. We also demonstrate, for the first time, fabrication and measurement of a free-standing thin Bi$_2$Se$_3$ membrane, which has low contact resistance to electrodes and a mobility of >~500 cm$^2$/Vs.

Recently, atomic membranes (AM) that are extracted from layered materials have become popular platforms for investigation of novel physical phenomena[1]. Some of the most studied materials are thin sheets of graphite[2], $Bi_2Se_3$ and $Bi_2Te_3$, which provide platforms for investigating massless Dirac fermions[3,4] and topological insulators[5-7]. Due to their two-dimensionality, they display a number of desirable characteristics such as gate tunable charge density and/or type, enhanced Coulomb interaction and coupling between local morphology and electronic properties[8,9]. As surface 2D electron systems, these membranes also enable optical and scanned probe measurements that are not possible in traditional semiconductor heterostructure devices.

Another significant advantage of these systems is that they can be easily coupled to special electrodes, such as superconductors or ferromagnets, potentially enabling experimental realization of some of the most fascinating predictions for these systems, such as specular Andreev Relfection[10], Majorana fermions[11,12] and spin Hall effect. Yet, interaction between AM and the substrate is known to be a significant impediment for the observation of such phenomena, since the substrate can locally dope the membranes, induce local corrugations and strains, and introduce scatterers such as charged impurities and surface phonons. Thus far, removing the substrate has proven to yield exceedingly high-quality devices[13,14], yielding novel phenomena such as Wigner crystallization and Mott insulating states in carbon nanotubes[15,16] and fractional quantum Hall effect in graphene[17,18].

To remove substrates, the most commonly adopted technique is acid etching, which dissolves the oxide layer underneath the device. However, this technique suffers from several drawbacks, including limitation of membrane and electrode materials to those that are stable in acid, and substrate to those that are not. For instance, superconducting and ferromagnetic materials cannot survive such a procedure and many of the much sought-after topological insulator materials, including $Bi_2Se_3$ and $Bi_2Te_3$, are also partially soluble in hydrofluoric acid that etches $SiO_2$. Here we report an innovative multilevel lithography technique to fabricate devices with free-standing AM extracted from layered materials. Employing only standard resists and developers for liftoff lithography, this technique can be applied to the vast majority of commonly used substrate, membrane and electrode materials, while imparting minimal damage to the device, which does not undergo any acid or reactive ion etching. Moreover, since the device is suspended above the substrate, the risk of gate leakage is minimized. Using this technique, we demonstrate the fabrication of freestanding graphene coupled to Ti/Al electrodes, with a number of widths and source-drain separations. We also report, for the first time, fabrication and measurement of suspended thin $Bi_2Se_3$ membranes, with estimated mobility of ~500 $cm^2$/Vs. In the future, this versatile technique can be used to explore, for instance, superconductivity and spintronics in high mobility graphene and $Bi_2Se_3$ samples, as well as in other layered materials.

This fabrication process is based on a method developed by us[19] to suspend local gates above graphene. The procedure, which consists of three electron beam lithography (EBL) steps, utilizes different exposure, developing and lifting off properties of different resists. Fig. 1 illustrates the entire procedure to create two suspended electrodes that contact a freestanding membrane. In the first step, we deposit and bake a layer of Lift-Off (LOR) resist onto a *p*-doped Si chip that is covered with a 310 nm-thick $SiO_2$ layer. AM sheets are directly exfoliated onto the LOR layer, and can be identified using atomic force microscope or color interference under an optical microscope (Fig. 1a). Subsequently, a bilayer of electron beam resists, MMA/PMMA, are spun and baked onto the sample, followed by exposure of alignment mark patterns and development in MIBK. These alignment marks are used for locating and aligning electrode patterns to the AM in the subsequent steps. We note that no metal deposition is necessary, as openings in the PMMA/MMA layer are sufficiently visible in the scanning electron microscope (SEM) for alignment, thus greatly simplifying the fabrication procedure.

In the next step, we use EBL to expose areas adjacent to the AM, and develop in MIBK/isopropyl alcohol(IPA) solution that dissolves only the exposed regions of the MMA/PMMA bilayer, but not LOR, so that two windows in the resist bilayer are created on either side of the AM. The exposed LOR within the windows is removed by developing in MF319, while the rest of the LOR layer remains intact (Fig. 1c). The end result of this step are four windows on the LOR/MMA/PMMA resist, which, after metallization, will form anchors on the substrate to connect to and support the two suspended electrodes.

In the third and final step of fabrication, we fabricate two suspended electrodes to contact the AM. To this end, we expose two long rectangular windows that lie directly on top of the AM and connect to the openings created in step 2(Fig. 1d). The chip is developed in MIBK/IPA to remove the exposed MMA/PMMA. We then perform metal deposition at 3 different angles (+45°, -45° and 0°) to ensure good contact at the sidewalls that attach the anchors to the suspended electrodes[20]. Finally, the samples are immersed in warm PG remover to remove all resist layers, and dried using a critical point dryer to prevent structural instability during the drying process. The end result is a sheet of atomic membrane "held up" by two partially suspended electrodes.

This powerful fabrication technique is versatile and robust. By tuning lithography parameters, we can produce free-standing electrodes that suspend layered materials with varying widths, lengths and heights. Using graphene as an example of AM, we fabricate a number of suspended devices with electrode separations ranging from 700 nm to 4 µm (Fig. 2), with a total suspended length as long as 40 µm (Fig. 2a inset). The graphene sheets usually exhibit no discernible structural deformation, though strain-induced ripples[21] have been occasionally observed (Fig. 2d). In these examples, graphene sheets are suspended at ~300 nm above the $SiO_2$ substrate, though this height can be easily adjusted from 50 nm to 3 µm by selecting different LOR solutions.

This acid-free fabrication technique is capable of producing devices with both long and short electrode separations. The latter geometry is particularly interesting for, *e.g.* realization of a ballistic graphene-based Josephson junction. Such a system has been predicted to exhibit several novel phenomena, such as specular andreev relection[22], chargeless transfer of spins[23] and thermopower[24], but has yet to be experimentally realized.

To demonstrate the viability of this fabrication procedure, we fabricate freestanding graphene devices with Ti/Al electrodes, and measure their transport characteristics using standard lock-in techniques at low temperature. In Fig. 2e-f we plot the conductance $G$ as a function of gate voltage $V_g$ for two different devices similar to the ones shown in Fig. 2c-d. Data displayed in Fig. 2e is obtained from a device with a source-drain separation of 1.7 µm, and graphene width 3.5µm. The red trace shows the device's initial $G(V_g)$ behavior immediately after fabrication. The relatively large conductance indicates small contact resistance; however, the poor response to gate and the absence of a Dirac point suggests that the device is highly doped. Such behavior is not uncommon for as-fabricated suspended graphene devices. After current annealing[13, 14, 25] at ~1.2 mA for 10 minutes, the device's behavior is significantly improved. As shown by the black trace, the Dirac point appears at $V_g$~0, and the $G(V_g)$ is symmetric with respect to the electron and hole branches. The device mobility is estimated to be ~3000 $cm^2$/Vs, which is reasonable and can be further optimized. The reproducibility of the results is demonstrated by similar behavior from a second device, with a mobility of 5500 cm$^2$/Vs (Fig. 2f). We also note that up to 80V in $V_g$ can be applied (or equivalently, up to 1.1x10$^{12}$ cm$^{-2}$ in induced charge density) without collapsing graphene, which is significantly higher than that applied to free-standing graphene devices fabricated via acid-release of SiO$_2$[3, 4], thus allowing access to regimes of high carrier density with rich many-body effects.

As a further demonstration of the versatility of this procedure, we fabricate suspended Bi$_2$Se$_3$ membranes, which is also a layered material. This topological insulator material is predicted to host a vast plethora of fascinating physical phenomena[7, 26]. Experimentally, the most illuminating results to date arise from data obtained from ARPES and STM measurements[27-29], while transport measurements have been limited[30, 31]. Here we demonstrate fabrication and measurement of suspended Bi$_2$Se$_3$ membranes, which have not been reported previously. The bulk samples are synthesized via Ca doping of single crystal Bi$_2$S$_3$ crystals[32]. Fig. 3a-b display SEM images of a completed Bi$_2$Se$_3$ device, which is ~65 nm thick. The current-voltage *(I-V)* characteristics of this device at 300K and 4K are both linear, with a resistance of 142 and 117 Ω, respectively (Fig. 3c). The linear *I-V* curves, in addition to the relatively small resistance that decreases with temperature, indicate low contact resistance and metallic conduction. Additionally, we also observed a small gate dependence: application of $\Delta V_g$=5V induces ~10 µS change in conductance, suggesting that, while much of the current is transported through the bulk, there is considerable surface conduction. Assuming all conduction is through the surface, the field effect mobility is estimated to be ~500 cm$^2$/Vs. We note that this is the lower-bound value for the surface mobility, which could be much higher if bulk conduction is taken into account[33]. In the future, we expect that device can be further optimized via improvement in material quality and reduction in membrane thickness.

In conclusion, we have developed a gentle and versatile multilevel lithography process to fabricate free-standing atomic membranes that are extracted from layered materials. Using this technique, we successfully suspended and performed measurements on atomically thin graphite and $Bi_2Se_3$ films that were coupled to Ti/Al electrodes. This technique provides a viable path towards the investigation of superconductivity and spintronics in high mobility graphene and $Bi_2Se_3$ samples, as well as in other layered materials.

We thank Desalegne Teweldebrhan for helpful discussions. This work was supported in part by ONR N00014-09-1-0724, the FENA Focus Center, UC Lab Fees Program and DARPA/DMEA under agreement number H94003-10-2-1003.

**Figure 1.** Schematics of fabrication process. (a). AM is exfoliated onto LOR (yellow) which rests on SiO$_2$/Si substrate (purple). (b). Bilayer MMA/PMMA (light gray) resists are deposited onto sample, and alignment cross marks are patterned by EBL. A second EBL is performed to expose regions indicated by the red arrows. (c). Developing in MIBK and MF319 removes both LOR and MMA/PMMA bilayer in the exposed regions. The final EBL step is performed to expose regions indicated by the red arrows. (d). Developing in MIBK removes *only* the MMA/PMMA resist bilayer in the exposed regions. (e) Metal deposition is performed at +45°, -45° and 0° using Ti/Al (dark gray). Samples are then immersed in warm PG remover and dried using a critical point dryer, leaving suspended electrodes that "hold" AM above the substrate.

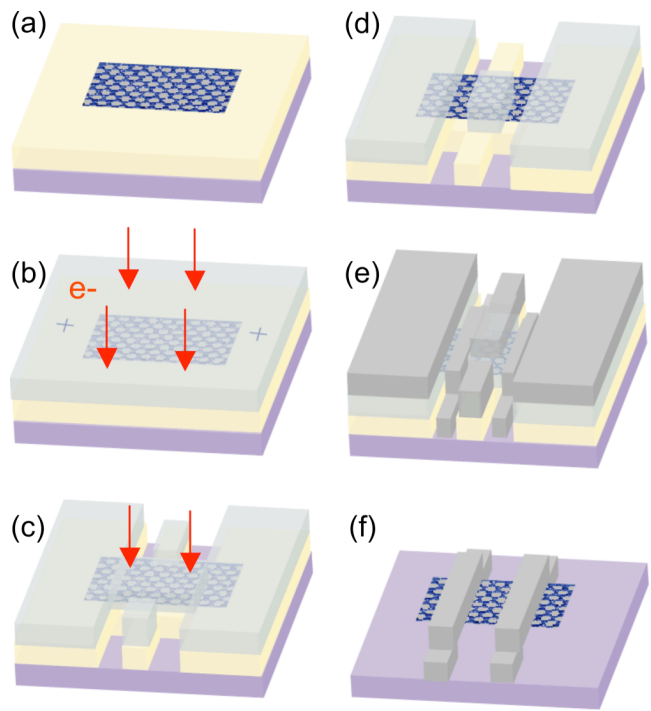

**Figure 2.** SEM images and transport data of suspended graphene devices. (a-d). SEM images of suspended graphene sheets with different widths and lengths. Scale bars: 1 μm. Inset: a 40-μm long graphene sheet suspended by several electrodes. (e-f). Device conductance as a function of gate voltage for two different suspended graphene devices. Red and black traces are for as-fabricated and current-annealed devices, respectively.

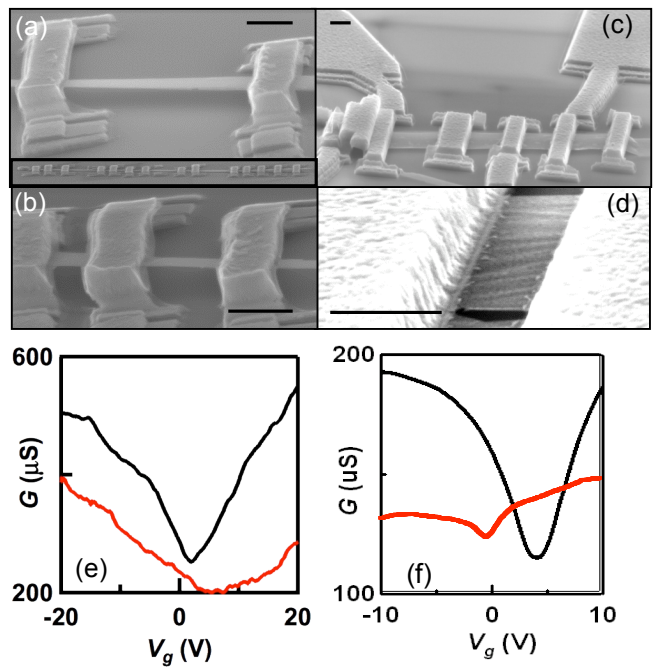

**Figure 3.** SEM images and transport data of a suspended $Bi_2Se_3$ device. (a-b) Top and angled view of a suspended $Bi_2Se_3$ device. Scale bar: 1 μm. The images are false-colored. (c). Current measured as a function of voltage bias at 300K (red) and 4K (blue).

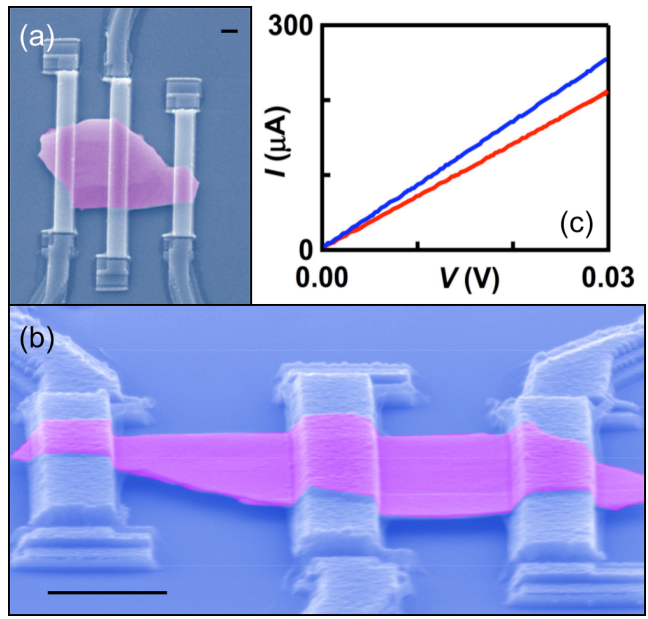